\documentclass[aps,prl,twocolumn,showpacs,superscriptaddress,floatfix]{revtex4-1}
\usepackage{graphicx}
\usepackage{amsmath,amssymb}
\usepackage[utf8]{inputenc}
\usepackage{natbib}
\usepackage{hyperref}
\usepackage{color}
\usepackage{IEEEtrantools}
\hypersetup{colorlinks=true,citecolor={blue},linkcolor={blue},urlcolor={blue}}
\usepackage{units}
\usepackage{float}
\newcommand{\Fig}[1]{Fig.\,\ref{#1}}
\newcommand{\fig}[1]{\,\ref{#1}}

\begin{document}
\title{A robust platform for engineering pure-quantum-state transitions in polariton condensates}

\date{\today}

\author{A. Askitopoulos}
\email[correspondence address: ]{Alexis.Askitopoulos@soton.ac.uk}
\affiliation{Faculty of Physical sciences and engineering, University of Southampton, Southampton, SO171BJ, United Kingdom}

\author{T.C.H. Liew}
\affiliation{School of Physical and Mathematical Sciences, Nanyang Technological University, 637371, Singapore}

\author{H. Ohadi}
\affiliation{Faculty of Physical sciences and engineering, University of Southampton, Southampton, SO171BJ, United Kingdom}
\email[Current address: ]{Department of Physics, Cavendish Laboratory, University of Cambridge, Cambridge CB3 0HE, United Kingdom}

\author{Z. Hatzopoulos}
\affiliation{Microelectronics Research Group, IESL-FORTH, P.O. Box 1527, 71110
Heraklion, Crete, Greece}
\affiliation{Department of Physics, University of Crete, 71003 Heraklion, Crete, Greece}

\author{P.G. Savvidis}
\affiliation{Microelectronics Research Group, IESL-FORTH, P.O. Box 1527, 71110
Heraklion, Crete, Greece}
\affiliation{Department of Materials Science and Technology, University of Crete, Crete, Greece}

\author{P.G. Lagoudakis}
%\email[correspondence address: ]{pavlos.lagoudakis@soton.ac.uk}
\affiliation{Faculty of Physical sciences and engineering, University of Southampton, Southampton, SO171BJ, United Kingdom}

\begin{abstract}
	We report on pure-quantum-state polariton condensates in optical annular traps. The study of the underlying mechanism reveals that the polariton wavefunction always coalesces in a single pure-quantum-state that, counter-intuitively, is always the uppermost confined state with the highest overlap to the exciton reservoir. The tunability of such states combined with the short polariton lifetime allows for ultrafast transitions between coherent mesoscopic wavefunctions of distinctly different symmetries rendering optically confined polariton condensates a promising platform for applications such as many-body quantum circuitry and continuous-variable quantum processing. 
\end{abstract}

\pacs{}
\maketitle

%\section{Introduction}
%Intro to MC polariton condensates $\rightarrow$ malleability of polariton mesoscopic wavefunctions $\rightarrow$ optical moulds $\rightarrow$ phase-locked polariton condensates/trapped condensates

Polaritons in semiconductor microcavities are light-matter bosonic quasi-particles formed by strong coupling of cavity photons and intra-cavity excitons ~\cite{deng2010}. Their excitonic part gives rise to strong interactions essential for fast thermalization and condensation, while their photonic part contributes to their very low effective mass ($~5 \times 10^{-5}m_{e}$) allowing for high temperature condensation ~\cite{roomtemperature}. Polariton condensates have been observed both under non-resonant optical excitation~\cite{BEC2006} and more recently under electrical injection of carriers~\cite{schneider_2013,solid_2013}. However, polaritons populate a two dimensional plane where a true Bose phase transition is theoretically possible only in the presence of a confining potential~\cite{berman_theory_2008} and this was first demonstrated with a stress induced trap~\cite{balili_bose-einstein_2007}. Unlike the weak atom-atom interactions in cold atomic Bose Einstein condensates (BEC), inter-particle interactions in a semiconductor microcavity are strong enough to substantially renormalise polariton self-energy, experimentally observed as a local blue-shift of the polariton spectrum. Variations of the polariton density in the plane of the cavity result in a potential landscape that can be externally controlled through real space modulation of the optical excitation beam. The malleability of the potential landscape can be used to imprint scattering centres~\cite{sanvitto_all-optical_2011} and devise polariton traps~\cite{Ring,chiral2014} and gates~\cite{gao_polariton_2012}. The dynamics of polariton condensates in externally modulated potential landscapes can lead to trapped states, standing polariton waves and phase-locking of remote condensates in non-trivial configurations~\cite{Ring,manni_spontaneous_2011,opticalSPT2013,rotating_2014,kalevich,ohadi_dissipative_2014}. Extensive control over mesoscopic polariton wavefunctions and their transitions between quantum states, coupled with the extensive propagation of polaritonic flows~\cite{sanvitto_all-optical_2011,schmutzler_all-optical_2015}, bares applications in quantum control, quantum circuits and on-chip quantum information processing~\cite{YamamotoReview_2014,qubits_2014}.

In this letter, we investigate the dynamics of pure quantum state transitions of polariton condensates under optical confinement. We utilise a ring-shaped, non-resonant optical excitation scheme to create a size-tunable annular potential trap. Under continuous wave excitation, we study the steady-state regime of trapping and condensate formation. We control the height of the potential trap by tuning the optical excitation density and observe that at coherence threshold, polaritons coalesce preferentially at the uppermost confined energy state that has the largest wavefunction overlap to the exciton reservoir that forms the trap barriers. To confirm that excited state polariton condensates are realised predominantly by polariton confinement in the optically induced potential trap, we study the transient dynamics of the formation mechanism. For this purpose, we change from continuous wave to pulsed excitation, while keeping all other parameters unaltered, and time-resolve the evolution of the spatial polariton state. Under pulsed excitation, the height of the potential barrier is transiently diminishing following the decay of the exciton reservoir. We observe that the mesoscopic polariton condensate switches between states, progressively transforming to the highest available confined energy state. The experimental observations are accurately reproduced using the extended Gross-Pitaevski equation.

\begin{figure*}
	\includegraphics[scale=0.35]{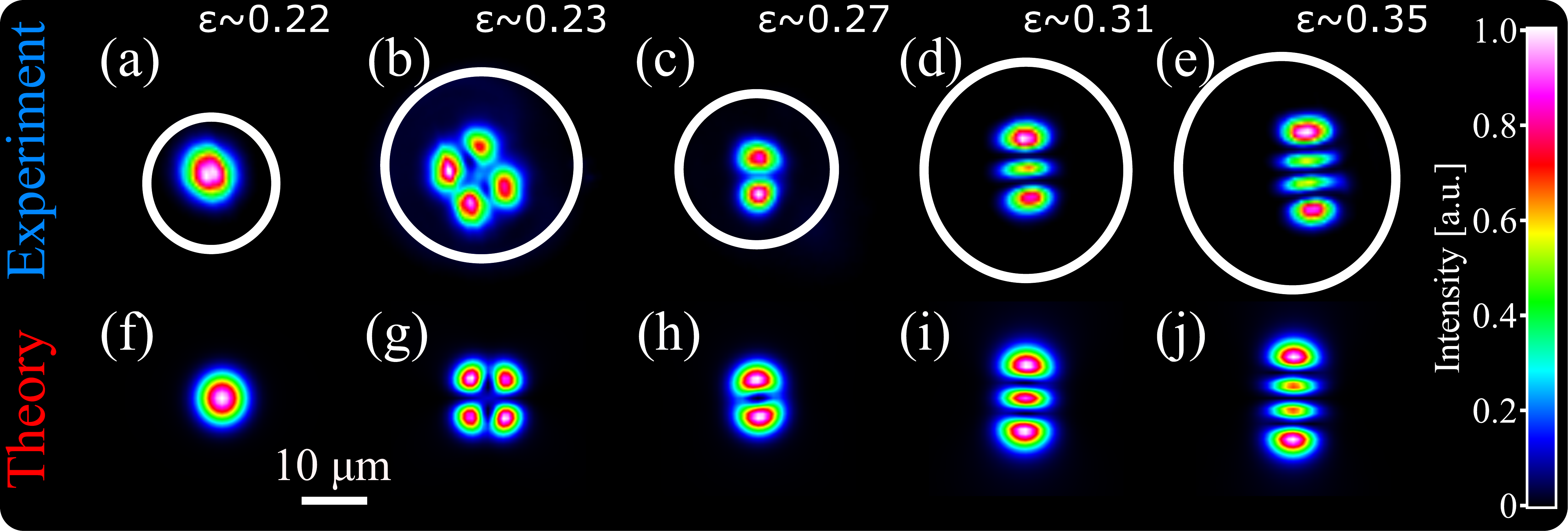}
	\centering %\vspace{-0.1cm}
	\caption{False color-scale experimental \textbf{(a)-(e)} and theoretical \textbf{(f)-(j)} states of polariton condensates.
	\textbf{(a),(f)} $\Psi_{00}$, \textbf{(b),(g)} $\Psi_{11}$, \textbf{(c),(h)} $\Psi_{01}$, 
	\textbf{(d),(i)} $\Psi_{02}$, \textbf{(e),(j)} $\Psi_{03}$. 
	$\epsilon$ denotes the ellipticity of individual configurations.}
	\label{fig:modes}
	\vspace{-0.4cm}
\end{figure*}

Non-ground state condensates of spatially-confined polaritons were previously observed in optical defect sites and in pillar microcavities, under Gaussian-shaped non-resonant optical excitation incident to the confinement area ~\cite{sanvitto2009,maragkou,nardin_2010}. While gain competition in thermodynamic equilibrium has been predicted to give rise to occupation of a single or several excited states~\cite{eastham2008,portolan2008}, in both cases, excited state condensates were shown to be driven by the dynamics of energy relaxation across the confined energy states resulting in multi-state condensation. In the case of ring-shaped excitation, two characteristically different regimes of polariton condensates have been realised. For ring radii comparable to the thermal de-Broglie wavelength a phase-locked standing-wave condensate co-localised to the excitation area was observed ~\cite{manni_spontaneous_2011}. For ring radii comparable to the polariton propagation length in the plane of the cavity, the excitation ring acted as a potential barrier and a Gaussian-shaped ground state polariton condensate was realized ~\cite{Ring}. 
Christofolini and co-workers examined the transition between phase-locked and trapped condensates using multiple-excitation spots and a ring-shaped excitation pattern~\cite{opticalSPT2013}. Despite earlier work by Manni et al~\cite{manni_spontaneous_2011}, the authors claimed that for ring-shaped pumps, no phase-locked state is geometrically possible, and that when the spacing between the pumps reduces, the trapped condensate collapses into a Gaussian-shaped ground state. Here, we show that under ring-shaped excitation, the formation of excited state condensates is driven by polariton confinement in the linear potentials and that the presence of non-ground polariton condensates does not necessitate asymmetries in the shape and/or power distribution of the ring excitation. The dependence of the state selection on the height of the trap's barrier and shape at threshold, provides a robust platform for engineering switches of mesoscopic multi-particle coherent states.

%\section{Experiment}

The experimental configuration that produces an annular beam of zero angular momenta consists of a double axicon arrangement. A variable telescope is used to control the radii of the excitation beam that we project on the sample. The excitation and detection configuration and the microcavity sample is described in ref.~\cite{Ring}. The microcavity is held in a cold finger cryostat operating at 6 K. We study the steady-state dynamics under non-resonant excitation at \unit[752]{nm} using a single mode quasi-continuous wave (CW) laser (2\% on-off ratio at 10kHz). The microcavity used in these experiments is a high Q factor ($>$ 15000) $5\lambda/2$ GaAs/AlGaAs microcavity with 4 triplets of \unit[10]{nm} GaAs quantum wells, with a Rabi splitting of \unit[9]{meV} and a cavity lifetime of \unit[7]{ps}, as described in ref.~\cite{tsotsis_lasing_2012}. All experiments were performed for a small negative detuning range of $\unit[-7]{meV}\leq$ d $\leq\unit[-5]{meV}$.

\begin{figure}[ht]
	\includegraphics[scale=0.45]{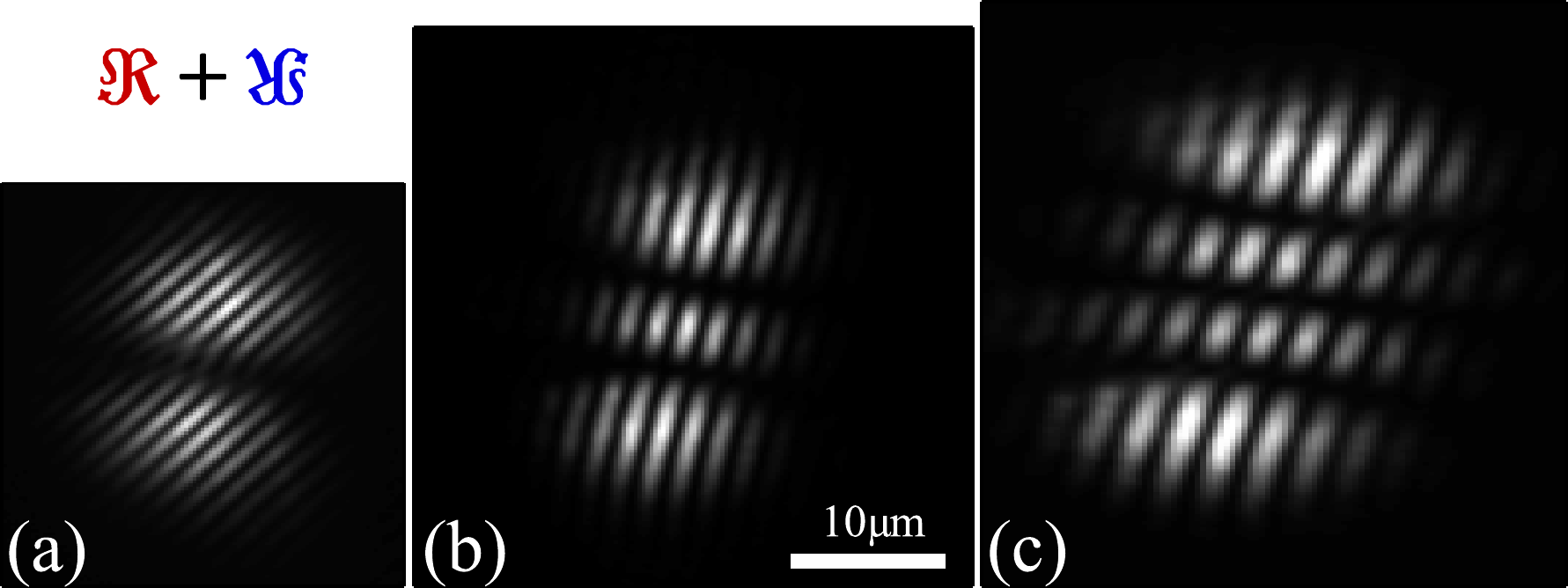}
	\centering \vspace{-0.25cm}
	\caption{\textbf{Interference patterns of trapped polariton condensates: (a)} $\Psi_{01}$, \textbf{(b)} $\Psi_{02}$, \textbf{(c)} $\Psi_{03}$. The interference patterns where obtained with a retro-reflector configuration.}
	\vspace{-0.2cm}
	\label{int}
\end{figure}

Figures \fig{fig:modes}a-e, show the spatial profile of mesoscopic wavefunctions for a range of excitation radii and asymmetries, characterised by the ellipticity and radius of the excitation ring, at the coherence threshold that defines the depth of the trap via the interactions in the reservoir. Theses states resemble the TEM modes of a harmonic oscillator and in what follows we will adapt their symbolism to annotate the state of the polariton wavefunction. For an excitation ring with a radius of $\sim$\unit[10]{$\mu$m} we observe a ground-state polariton condensate (\Fig{fig:modes}a), as in ref.~\cite{Ring}, which remains in the ground-state as long as the long axis of the asymmetric excitation does not exceed $\sim$\unit[10]{$\mu$m}. For larger excitation ring radius ($\sim$\unit[17]{$\mu$m}) and similar ellipticity as in \Fig{fig:modes}a ($\epsilon=0.22$) at coherence threshold we observe that polaritons coalesce at a higher excited state ($\psi_{11}$) as shown in \Fig{fig:modes}b. We note that the symmetry of the excited state wavefunction is robust to small asymmetries in the excitation ring ($0<\epsilon<0.23$) and the transition from ground to non-ground polariton condensates is predominantly dependent on the radius of the ring. Increasing the ring radius and the asymmetry of the excitation it is possible to observe excited state polariton condensates as shown in \Fig{fig:modes}c-e. On top of each panel we have annotated the ellipticity of the excitation ring. Interferometric measurements of excited states $\psi_{01},\psi_{02},\psi_{03}$ confirm that these are coherent mesoscopic wavefunctions of extended condensates (\Fig{int}a-c). 

\begin{figure}[ht]
	\includegraphics[scale=0.25]{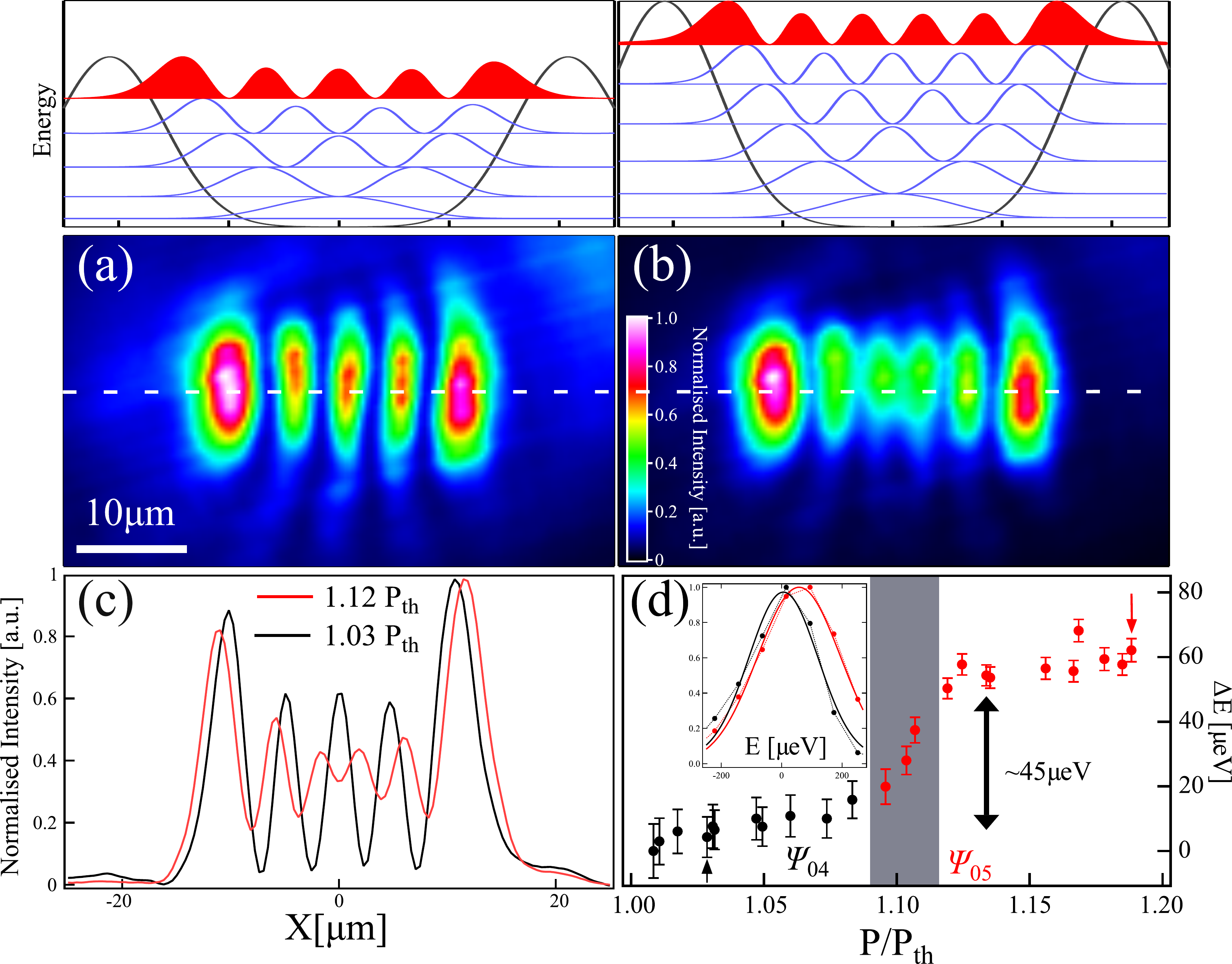}
	\centering 
	\caption{\textbf{Evolution of $\Psi_{04}$ for increasing excitation density. Bottom panel: (a)}  $\Psi_{04}$ at P=1.03$P_{th}$. \textbf{(b)} Subsequent increase of the power results in the appearance of $\Psi_{05}$. Top panel: Schematic representation of the confined energy states for two different barrier heights. \textbf{(c)} Profiles of the wavefunction for different excitation densities extracted along the dashed white lines at \textbf{(a)} and \textbf{(b)}. \textbf{(d)} Corresponding energy difference with respect to the energy at coherence threshold for increasing excitation power normalised at the coherence threshold power $P_{th}$. Inset in \textbf{(d)} shows the spectra of the points denoted by the arrows}
	\label{Ptran}
	\vspace{-0.25cm}
\end{figure}

We investigate the dependence of the quantum state selectivity on the barrier height by varying the non-resonant excitation density of a geometrically fixed, ring-shaped, asymmetric excitation profile. We use an excitation ring of radius $\sim$16 $\mu m$ and $\epsilon=0.27$ that at coherence threshold produces the $\Psi_{04}$ polariton state as shown in Fig.\ref{Ptran}a. By increasing the excitation density above the coherence threshold, while keeping all other parameters the same, we observe the transition from $\Psi_{04}$ to $\Psi_{05}$ (Fig. \ref{Ptran}b). The order of the latter state is clearly revealed in Fig.\ref{Ptran}c, where we plot the normalised spatial profiles along the white dashed lines of the real space intensity images of Fig.\ref{Ptran}a,b. Fig.\ref{Ptran}c shows the presence of an extra node at the higher excitation density indicative of $\Psi_{05}$. In Fig. \ref{Ptran}d we plot the energy shift of the condensate in the transition from $\Psi_{04}$ to $\Psi_{05}$ with respect to its energy at the coherence threshold ($\Delta(E_P-E_{P_{th}})$). A sharp increase of the energy shift ($\sim 45\mu eV$) is observed in Fig.\ref{Ptran}d at $P\sim 1.12 P_{th}$. Within the grey stripe intensity fluctuations of the excitation beam artificially blur the two states. The top panels in Fig.\ref{Ptran}a,b depict the calculated energy levels for the trap shape and the corresponding probability density of the confined states. In both panels, the red-filled probability density corresponds to the occupied state. It is worth noting here the greater overlap of the probability density of the highest energy level ($\Psi_{04}$ in \ref{Ptran}(a) and $\Psi_{05}$ \ref{Ptran}(b)) with the reservoir compared to the lowest energy levels. Evidently, with increasing the barrier height a polariton condensate is realised at the next confined energy level as a pure-quantum-state that can be singularly described by the principal quantum number $n$ $(\Psi_{0,n+1})$.  

\begin{figure}[ht]
	\includegraphics[scale=0.29]{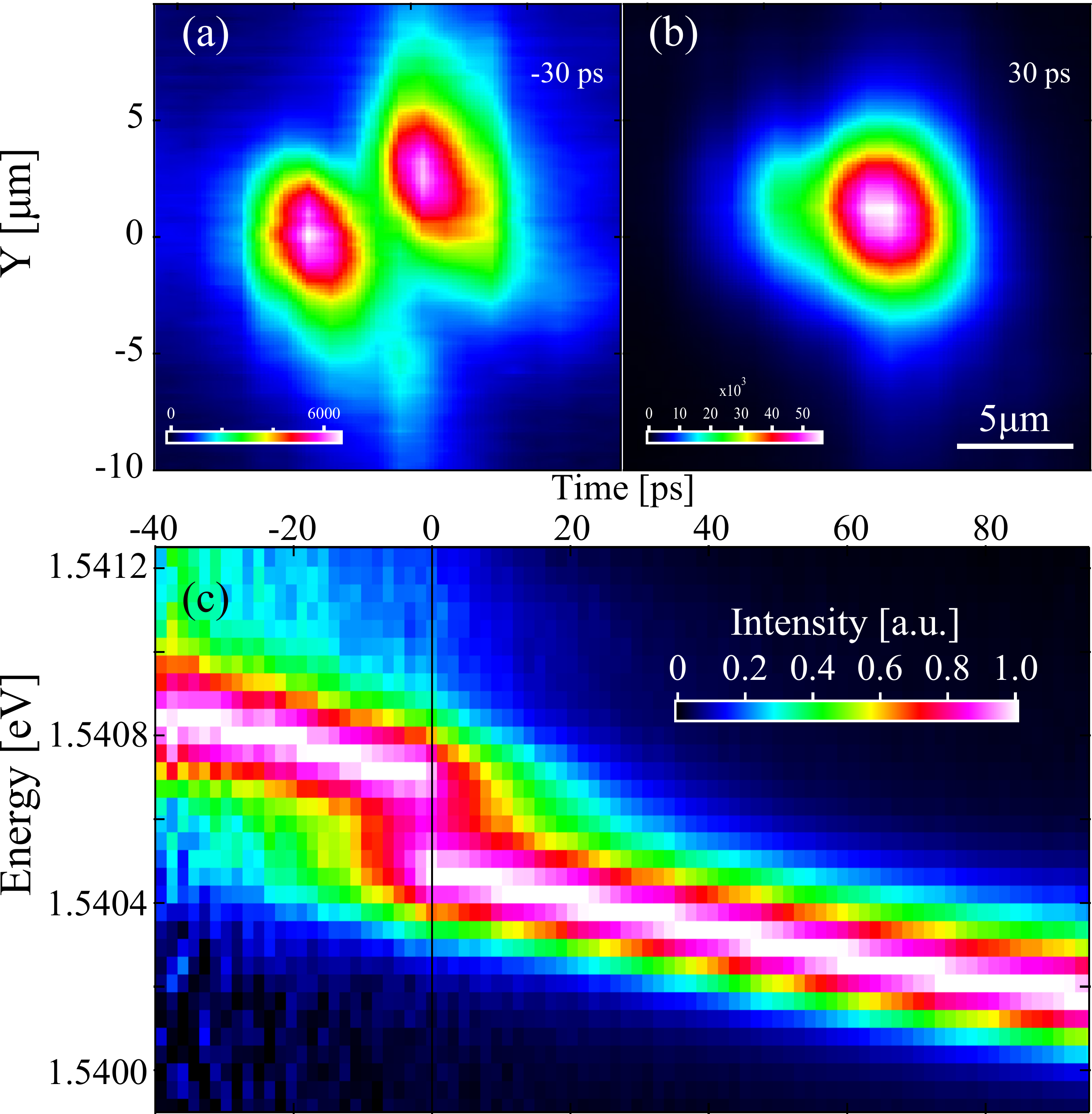}
	\centering \vspace{-0.35cm}
	\caption{\textbf{False colour-scale real space tomographic frames in the time domain. (a)} $\Psi{01}$ state at -30ps and \textbf{(b)} subsequent 
		transition to $\Psi_{00}$ at 30ps. \textbf{(c)} Intensity normalised time evolution of the emission energy showing the characteristic energy jump upon the transition threshold.}
	\label{Ttran}
	\vspace{-0.25cm}
\end{figure}

We explore the robustness of the formation of pure-quantum-states on density fluctuations in the exciton reservoir, by extending our study from the excitation density dependent switching between successive states in the dynamic equilibrium regime, to transitions in the time domain under non-resonant pulsed excitation. We use a ring-shaped non-resonant $200$ femtosecond pulse at \unit[755]{nm} with $\sim$\unit[11]{$\mu$m} radius of the major axis and $\epsilon= 0.3$ at $\sim$1.6$P_{th}$. We record the spatio-temporal dynamics of the emission and observe the formation of the $\Psi_{01}$ polariton state and its transition to $\Psi_{00}$~\cite{supp}. We set the transition point to define the zero time frame for the rest of our analysis. Figure \ref{Ttran}a shows a snapshot of the $\Psi_{01}$ state at \unit[-30]{ps}. At later times, the two lobes of the $\Psi_{01}$ state appear to move closer together and the condensate rapidly transforms to the ground polariton state ($\Psi_{00}$) of Fig.\ref{Ttran}b. The decrease of the density in the  barriers in the time-domain results in a shallower trap in which the $\Psi_{01}$ state is no longer confined, leading to a polariton condensate at the next available state, here the ground state $\Psi_{00}$. We spectrally and time resolve the decay of emission at normal incidence with an angular width corresponding to $|k|\leq 1.4 \mu$m and observe a sharp energy shift from $\Psi_{01}$ to $\Psi_{00}$ as shown in Figure \ref{Ttran}c. This dynamic transition further illustrates that under optical confinement a polariton condensate spontaneously occurs at a higher confined state as defined by the barrier height of the trap and that the transition to the ground state is hindered solely by the existence of higher energy levels. 
%\vspace{-0.4cm}

\begin{figure*}
	\vspace{-0.1cm}
	\includegraphics[scale=0.29]{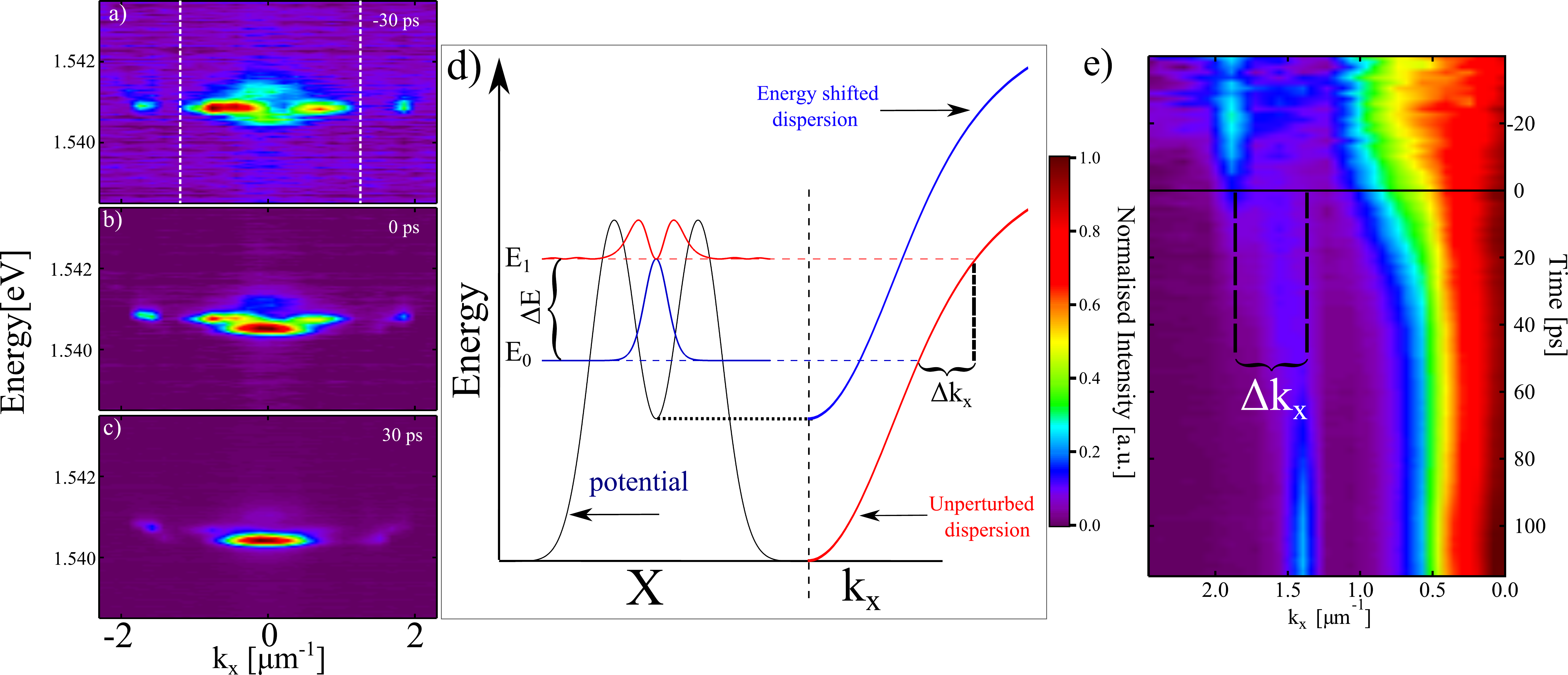}
	\centering \vspace{-0.15cm}
	\caption{ Polariton dispersion at -30ps a), 0ps, b) and 30ps c). White dotted lines in a) denote the integrated area from which Fig.\fig{Ttran}c was extracted. d) Schematic representation of the momentum acquired by polaritons tunnelling outside of the potential trap (potential and dispersion energy not in scale). e) Intensity normalised time evolution of $k_x$ showing the characteristic $\Delta k_x$ jump of the tunnelling mode.}
	\label{tunel}
	\vspace{-0.25cm}
\end{figure*}

The time resolved dispersion images from which the energy evolution of the system was extracted (Fig.\fig{Ttran}c) are presented in Fig.\fig{tunel}a-c.
The appearance of the $\Psi_{01}$ mode is accompanied by a distinct doublet mode in the dispersion (Fig.~\fig{tunel}a), which corresponds to the counter-propagating components of the standing wave~\cite{rotating_2014}. As the barrier dynamically decays and the $\Psi_{00}$ mode is switched on as previously discussed, it quickly overtakes $\Psi_{01}$ in intensity at $\sim$\unit{0}{ps}. The first excited state quickly dissipates after this point with the polariton lifetime and the dispersion is dominated by the emission of the trap ground state. Interestingly, Fig.\fig{tunel}a-c also reveal distinct satellite modes at the same energy of the confined modes but for greater in plane wave-vector. For quantum states in traps with a finite barrier width, coherent tunnelling modes are a characteristic feature. Moreover, in our system these modes will be accelerated by the potential landscape outside the trap eventually acquiring momentum characteristic of the difference between the energy level in the trap and of the low-density polariton dispersion of the system outside the excitation region (Fig.\fig{tunel}d). From this description it becomes clear that the tunnelling modes are expected to be at the same energy but with higher momentum, as observed in Fig.\fig{tunel}a-c. 

Integrating the time-dispersion images over energy, while intensity normalising for every time frame, we compile the time evolution of k$_x$ (Fig.\fig{tunel}e). This analysis reveals the expected $\Delta$k$_x$ difference of the tunnelling modes of the two states. Intuitively, the relative (to the trapped state) intensity of the $\Psi_{01}$ tunnelling mode at the transition is substantial, as the width of the barrier goes to zero at this energy level. In contrast to the tunnelling amplitude of the ground state that is effectively suppressed as the potential width at the $\Psi_{01}$ energy level is still significant. Nevertheless, following the dynamic dissipation of the barrier, due to the decay of particles as well as draining of the reservoir by the condensate, we observe a continuous increase of the relative intensity of the tunnelling amplitude of the ground state at k$_x\sim \unit{1.4}{\mu m ^{-1}}$. The observation of a strong tunnelling component from the E$_{01}$ energy just before the transition, verifies that the barrier width for this level is indeed minimal and that $\Psi_{01}$ is close to the rim of the trap barrier, further corroborating our interpretation.
The system can be theoretically modelled with a non-linear Schr\"{o}dinger equation, namely the Gross-Pitaevski equation.
Simulations with the Gross-Pitaevski equation with a potential similar to the one from the experimental measurements in our system 
qualitatively reproduce the states recorded experimentally. Using a potential $V(r)$ that consists of the exciton-exciton interactions in 
the reservoir, that blue-shift the polariton energy levels, and of the polariton-polariton interactions in the condensate,the Hamiltonian of the system is: 
\begin{IEEEeqnarray}{l}
H(r) = T+V(r) \\
V(r)  =  V_{r}(r)+V_{c}(r)  \\
 V_{r}(r)=N_{r}U_{ex-ex} f_{r}(r)
\end{IEEEeqnarray}
where $N_{r}$ is the density of excitons in the reservoir, $U_{ex-ex}$ the exciton-exciton interaction 
strength, $f_r(\mathbf{r})$ is the spatial distribution of the exciton 
reservoir taking into account exciton diffusion beyond the pump spot and $V_c=U_{pol-pol}|\psi_n(\mathbf{r})|^2$ where $U_{pol-pol}$ the polariton-polariton interaction strength and $\psi(\mathbf{r})$ the condensate wavefunction. In addition to kinetic and potential energy terms in the above Hamiltonian, to account for 
polariton spatial dynamics, a generalization of the extended Gross-Pitaevskii equation is required to include incoherent pumping and 
decay~\cite{Wouters2007}. In continuous wave experiments one expects the excitation of a steady state of hot excitons with the spatial profile set by the optical pumping extended by exciton diffusion. One can then make use of the Landau-Ginzburg approach for describing the dynamics of the 2D polariton wavefunction~\cite{Keeling2008}:
\begin{align}
i\hbar\frac{d\psi(\mathbf{r},t)}{dt}&=\left[-\frac{\hbar^2\hat{\nabla}^2}{2m_P}+\left(U_{pol-pol}-i\Gamma_\mathrm{NL}\right)|\psi(\mathbf{r},t)|^2\right.\notag\\
&\hspace{10mm}\left.+\left(U_{pol-ex}+ir\right)N_rf_r(\mathbf{r})-\frac{i\Gamma}{2}\right]\psi(\mathbf{r},t)\notag\\
&\hspace{10mm}+i\hbar\mathfrak{R}\left[\psi(\mathbf{r},t)\right].\label{eq:GP}
\end{align}
Here $m_P$ is the polariton effective mass and $f_r(\mathbf{r})$ describes the 2D spatial distribution of $N_r$ excitons. The condensation rate $r$ describes the gain of polaritons in the presence of the exciton reservoir. The polaritons experience both a linear decay $\Gamma$ and non-linear loss $\Gamma_{NL}$, which represents the scattering of polaritons out of the condensate when its density is high~\cite{Keeling2008}. The final term in Eq.~\ref{eq:GP} represents a phenomenological energy relaxation~\cite{Wouters2012} in the system, which can play an important role when non-ground state polaritons interact with a potential gradient~\cite{Wouters2010,Wertz2012,Anton2012}:
\begin{equation}
\mathfrak{R}[\psi(x,t)]=-\lambda N_rf_r(\mathbf{r})\left(\hat{E}_\mathrm{LP}-\mu(\mathbf{r},t)\right)\psi(x,t).\label{eq:relax}
\end{equation}
where $\lambda$ determines the strength of energy relaxation~\cite{Wouters2012,Wertz2012} and $\mu(\mathbf{r},t)$ is a local effective chemical potential that conserves the polariton population~\cite{Wouters2012}. Kinetic energy relaxation of this form was derived with a variety of methods~\cite{Solnyshkov2014,Sieberer2014} and offers a simple model for the qualitative description of our experiment. We note however that this model does not distinguish between different mechanisms of energy relaxation, which may have different power dependences~\cite{Haug2014}.

Fixing $N_rf_r(\mathbf{r})$ to represent a ring shaped excitation (with slight asymmetry), the numerical solution of Eq.~\ref{eq:GP} gives the steady state intensity profiles shown in \Fig{fig:modes}f-j. Different configurations are accessed by varying the spatial distribution ($f_r(\mathbf{r})$) and population ($N_r$) of hot excitons, as in the experiment~\cite{TimV}. The simulations support that excited state condensation occurs preferentially at the uppermost confined energy state.

Although it cannot be explicitly verified that there is no available state in the trap above the condensate energy level, since the polariton potential landscape is not directly measurable, the evidence presented from the steady state switching, the transient study including the dynamic behaviour of the tunnelling components of the system, as well as the theoretical simulations and the calculations for the condensate reservoir overlap~\cite{supp} strongly supports our interpretation that polaritons condense in the highest available energy state within the optical trap.

In conclusion, we have investigated the dynamics of polariton condensates under optical confinement and observed that, in contrast to previously reported excited state condensation in defect traps and pillar structures, injection of polaritons from the trap barriers leads to the formation of a pure quantum-confined state with a mesoscopic coherent wavefunction above condensation threshold. This behaviour is in agreement with theoretical expectations for a true Bose condensate that is anticipated to resist multi-mode behaviour~\cite{combescot_stability_2008,nelsen_dissipationless_2013} in the presence of inter-particle interactions.  Moreover we revealed that the state selectivity of this system strongly depends on the geometric properties of the trap and have demonstrated a highly controllable switching between successive mesoscopic coherent quantum confined states, in the dynamic equilibrium regime and in the time domain. These results highlight the capability of tailoring and manipulating on-chip pure-quantum-states in semiconductor microcavities that can facilitate the implementation of polariton bosonic cascade lasers~\cite{liew_proposal_2013}. Taking into account that the extensive propagation~\cite{nelsen_dissipationless_2013} as well as the susceptibility of the polaritonic flow to the potential landscape~\cite{nguyen_realization_2013} has been widely demonstrated, these results also indicate the potential of engineering confined condensate lattices, coupled by their respective tunnelling amplitudes. Moreover the coupling strength in this architecture can be finely tuned by controlling the barrier height enabling the emergence of applications such as many-body quantum circuitry and quantum simulators.

%\section{Acknowledgements}
P. S. acknowledges funding from Greek GSRT programm APOLLO. A. A. acknowledges useful discussions with W. Langbein and S. Portolan.

%\bibliography{Bibliography}

\begin{thebibliography}{40}%
	\makeatletter
	\providecommand \@ifxundefined [1]{%
		\@ifx{#1\undefined}
	}%
	\providecommand \@ifnum [1]{%
		\ifnum #1\expandafter \@firstoftwo
		\else \expandafter \@secondoftwo
		\fi
	}%
	\providecommand \@ifx [1]{%
		\ifx #1\expandafter \@firstoftwo
		\else \expandafter \@secondoftwo
		\fi
	}%
	\providecommand \natexlab [1]{#1}%
	\providecommand \enquote  [1]{``#1''}%
	\providecommand \bibnamefont  [1]{#1}%
	\providecommand \bibfnamefont [1]{#1}%
	\providecommand \citenamefont [1]{#1}%
	\providecommand \href@noop [0]{\@secondoftwo}%
	\providecommand \href [0]{\begingroup \@sanitize@url \@href}%
	\providecommand \@href[1]{\@@startlink{#1}\@@href}%
	\providecommand \@@href[1]{\endgroup#1\@@endlink}%
	\providecommand \@sanitize@url [0]{\catcode `\\12\catcode `\$12\catcode
		`\&12\catcode `\#12\catcode `\^12\catcode `\_12\catcode `\%12\relax}%
	\providecommand \@@startlink[1]{}%
	\providecommand \@@endlink[0]{}%
	\providecommand \url  [0]{\begingroup\@sanitize@url \@url }%
	\providecommand \@url [1]{\endgroup\@href {#1}{\urlprefix }}%
	\providecommand \urlprefix  [0]{URL }%
	\providecommand \Eprint [0]{\href }%
	\providecommand \doibase [0]{http://dx.doi.org/}%
	\providecommand \selectlanguage [0]{\@gobble}%
	\providecommand \bibinfo  [0]{\@secondoftwo}%
	\providecommand \bibfield  [0]{\@secondoftwo}%
	\providecommand \translation [1]{[#1]}%
	\providecommand \BibitemOpen [0]{}%
	\providecommand \bibitemStop [0]{}%
	\providecommand \bibitemNoStop [0]{.\EOS\space}%
	\providecommand \EOS [0]{\spacefactor3000\relax}%
	\providecommand \BibitemShut  [1]{\csname bibitem#1\endcsname}%
	\let\auto@bib@innerbib\@empty
	%</preamble>
	\bibitem [{\citenamefont {Deng}\ \emph {et~al.}(2010)\citenamefont {Deng},
		\citenamefont {Haug},\ and\ \citenamefont {Yamamoto}}]{deng2010}%
	\BibitemOpen
	\bibfield  {author} {\bibinfo {author} {\bibfnamefont {H.}~\bibnamefont
			{Deng}}, \bibinfo {author} {\bibfnamefont {H.}~\bibnamefont {Haug}}, \ and\
		\bibinfo {author} {\bibfnamefont {Y.}~\bibnamefont {Yamamoto}},\ }\href
	{\doibase 10.1103/RevModPhys.82.1489} {\bibfield  {journal} {\bibinfo
			{journal} {Reviews of Modern Physics}\ }\textbf {\bibinfo {volume} {82}},\
		\bibinfo {pages} {1489} (\bibinfo {year} {2010})}\BibitemShut {NoStop}%
	\bibitem [{\citenamefont {Christopoulos}\ \emph {et~al.}(2007)\citenamefont
		{Christopoulos}, \citenamefont {von Högersthal}, \citenamefont {Grundy},
		\citenamefont {Lagoudakis}, \citenamefont {Kavokin}, \citenamefont
		{Baumberg}, \citenamefont {Christmann}, \citenamefont {Butté}, \citenamefont
		{Feltin}, \citenamefont {Carlin},\ and\ \citenamefont
		{Grandjean}}]{roomtemperature}%
	\BibitemOpen
	\bibfield  {author} {\bibinfo {author} {\bibfnamefont {S.}~\bibnamefont
			{Christopoulos}}, \bibinfo {author} {\bibfnamefont {G.~B.~H.}\ \bibnamefont
			{von Högersthal}}, \bibinfo {author} {\bibfnamefont {A.~J.~D.}\ \bibnamefont
			{Grundy}}, \bibinfo {author} {\bibfnamefont {P.~G.}\ \bibnamefont
			{Lagoudakis}}, \bibinfo {author} {\bibfnamefont {A.~V.}\ \bibnamefont
			{Kavokin}}, \bibinfo {author} {\bibfnamefont {J.~J.}\ \bibnamefont
			{Baumberg}}, \bibinfo {author} {\bibfnamefont {G.}~\bibnamefont
			{Christmann}}, \bibinfo {author} {\bibfnamefont {R.}~\bibnamefont {Butté}},
		\bibinfo {author} {\bibfnamefont {E.}~\bibnamefont {Feltin}}, \bibinfo
		{author} {\bibfnamefont {J.-F.}\ \bibnamefont {Carlin}}, \ and\ \bibinfo
		{author} {\bibfnamefont {N.}~\bibnamefont {Grandjean}},\ }\href {\doibase
		10.1103/PhysRevLett.98.126405} {\bibfield  {journal} {\bibinfo  {journal}
			{Physical Review Letters}\ }\textbf {\bibinfo {volume} {98}},\ \bibinfo
		{pages} {126405} (\bibinfo {year} {2007})}\BibitemShut {NoStop}%
	\bibitem [{\citenamefont {Kasprzak}\ \emph {et~al.}(2006)\citenamefont
		{Kasprzak}, \citenamefont {Richard}, \citenamefont {Kundermann},
		\citenamefont {Baas}, \citenamefont {Jeambrun}, \citenamefont {Keeling},
		\citenamefont {Marchetti}, \citenamefont {Szymańska}, \citenamefont
		{André}, \citenamefont {Staehli}, \citenamefont {Savona}, \citenamefont
		{Littlewood}, \citenamefont {Deveaud},\ and\ \citenamefont {Dang}}]{BEC2006}%
	\BibitemOpen
	\bibfield  {author} {\bibinfo {author} {\bibfnamefont {J.}~\bibnamefont
			{Kasprzak}}, \bibinfo {author} {\bibfnamefont {M.}~\bibnamefont {Richard}},
		\bibinfo {author} {\bibfnamefont {S.}~\bibnamefont {Kundermann}}, \bibinfo
		{author} {\bibfnamefont {A.}~\bibnamefont {Baas}}, \bibinfo {author}
		{\bibfnamefont {P.}~\bibnamefont {Jeambrun}}, \bibinfo {author}
		{\bibfnamefont {J.~M.~J.}\ \bibnamefont {Keeling}}, \bibinfo {author}
		{\bibfnamefont {F.~M.}\ \bibnamefont {Marchetti}}, \bibinfo {author}
		{\bibfnamefont {M.~H.}\ \bibnamefont {Szymańska}}, \bibinfo {author}
		{\bibfnamefont {R.}~\bibnamefont {André}}, \bibinfo {author} {\bibfnamefont
			{J.~L.}\ \bibnamefont {Staehli}}, \bibinfo {author} {\bibfnamefont
			{V.}~\bibnamefont {Savona}}, \bibinfo {author} {\bibfnamefont {P.~B.}\
			\bibnamefont {Littlewood}}, \bibinfo {author} {\bibfnamefont
			{B.}~\bibnamefont {Deveaud}}, \ and\ \bibinfo {author} {\bibfnamefont
			{L.~S.}\ \bibnamefont {Dang}},\ }\href {\doibase 10.1038/nature05131}
	{\bibfield  {journal} {\bibinfo  {journal} {Nature}\ }\textbf {\bibinfo
			{volume} {443}},\ \bibinfo {pages} {409} (\bibinfo {year}
		{2006})}\BibitemShut {NoStop}%
	\bibitem [{\citenamefont {Schneider}\ \emph {et~al.}(2013)\citenamefont
		{Schneider}, \citenamefont {Rahimi-Iman}, \citenamefont {Kim}, \citenamefont
		{Fischer}, \citenamefont {Savenko}, \citenamefont {Amthor}, \citenamefont
		{Lermer}, \citenamefont {Wolf}, \citenamefont {Worschech}, \citenamefont
		{Kulakovskii}, \citenamefont {Shelykh}, \citenamefont {Kamp}, \citenamefont
		{Reitzenstein}, \citenamefont {Forchel}, \citenamefont {Yamamoto},\ and\
		\citenamefont {Höfling}}]{schneider_2013}%
	\BibitemOpen
	\bibfield  {author} {\bibinfo {author} {\bibfnamefont {C.}~\bibnamefont
			{Schneider}}, \bibinfo {author} {\bibfnamefont {A.}~\bibnamefont
			{Rahimi-Iman}}, \bibinfo {author} {\bibfnamefont {N.~Y.}\ \bibnamefont
			{Kim}}, \bibinfo {author} {\bibfnamefont {J.}~\bibnamefont {Fischer}},
		\bibinfo {author} {\bibfnamefont {I.~G.}\ \bibnamefont {Savenko}}, \bibinfo
		{author} {\bibfnamefont {M.}~\bibnamefont {Amthor}}, \bibinfo {author}
		{\bibfnamefont {M.}~\bibnamefont {Lermer}}, \bibinfo {author} {\bibfnamefont
			{A.}~\bibnamefont {Wolf}}, \bibinfo {author} {\bibfnamefont {L.}~\bibnamefont
			{Worschech}}, \bibinfo {author} {\bibfnamefont {V.~D.}\ \bibnamefont
			{Kulakovskii}}, \bibinfo {author} {\bibfnamefont {I.~A.}\ \bibnamefont
			{Shelykh}}, \bibinfo {author} {\bibfnamefont {M.}~\bibnamefont {Kamp}},
		\bibinfo {author} {\bibfnamefont {S.}~\bibnamefont {Reitzenstein}}, \bibinfo
		{author} {\bibfnamefont {A.}~\bibnamefont {Forchel}}, \bibinfo {author}
		{\bibfnamefont {Y.}~\bibnamefont {Yamamoto}}, \ and\ \bibinfo {author}
		{\bibfnamefont {S.}~\bibnamefont {Höfling}},\ }\href {\doibase
		10.1038/nature12036} {\bibfield  {journal} {\bibinfo  {journal} {Nature}\
		}\textbf {\bibinfo {volume} {497}},\ \bibinfo {pages} {348} (\bibinfo {year}
		{2013})}\BibitemShut {NoStop}%
	\bibitem [{\citenamefont {Bhattacharya}\ \emph {et~al.}(2013)\citenamefont
		{Bhattacharya}, \citenamefont {Xiao}, \citenamefont {Das}, \citenamefont
		{Bhowmick},\ and\ \citenamefont {Heo}}]{solid_2013}%
	\BibitemOpen
	\bibfield  {author} {\bibinfo {author} {\bibfnamefont {P.}~\bibnamefont
			{Bhattacharya}}, \bibinfo {author} {\bibfnamefont {B.}~\bibnamefont {Xiao}},
		\bibinfo {author} {\bibfnamefont {A.}~\bibnamefont {Das}}, \bibinfo {author}
		{\bibfnamefont {S.}~\bibnamefont {Bhowmick}}, \ and\ \bibinfo {author}
		{\bibfnamefont {J.}~\bibnamefont {Heo}},\ }\href {\doibase
		10.1103/PhysRevLett.110.206403} {\bibfield  {journal} {\bibinfo  {journal}
			{Physical Review Letters}\ }\textbf {\bibinfo {volume} {110}},\ \bibinfo
		{pages} {206403} (\bibinfo {year} {2013})}\BibitemShut {NoStop}%
	\bibitem [{\citenamefont {Berman}\ \emph {et~al.}(2008)\citenamefont {Berman},
		\citenamefont {Lozovik},\ and\ \citenamefont {Snoke}}]{berman_theory_2008}%
	\BibitemOpen
	\bibfield  {author} {\bibinfo {author} {\bibfnamefont {O.~L.}\ \bibnamefont
			{Berman}}, \bibinfo {author} {\bibfnamefont {Y.~E.}\ \bibnamefont {Lozovik}},
		\ and\ \bibinfo {author} {\bibfnamefont {D.~W.}\ \bibnamefont {Snoke}},\
	}\href {\doibase 10.1103/PhysRevB.77.155317} {\bibfield  {journal} {\bibinfo
		{journal} {Physical Review B}\ }\textbf {\bibinfo {volume} {77}},\ \bibinfo
	{pages} {155317} (\bibinfo {year} {2008})}\BibitemShut {NoStop}%
\bibitem [{\citenamefont {Balili}\ \emph {et~al.}(2007)\citenamefont {Balili},
	\citenamefont {Hartwell}, \citenamefont {Snoke}, \citenamefont {Pfeiffer},\
	and\ \citenamefont {West}}]{balili_bose-einstein_2007}%
\BibitemOpen
\bibfield  {author} {\bibinfo {author} {\bibfnamefont {R.}~\bibnamefont
		{Balili}}, \bibinfo {author} {\bibfnamefont {V.}~\bibnamefont {Hartwell}},
	\bibinfo {author} {\bibfnamefont {D.}~\bibnamefont {Snoke}}, \bibinfo
	{author} {\bibfnamefont {L.}~\bibnamefont {Pfeiffer}}, \ and\ \bibinfo
	{author} {\bibfnamefont {K.}~\bibnamefont {West}},\ }\href {\doibase
	10.1126/science.1140990} {\bibfield  {journal} {\bibinfo  {journal}
		{Science}\ }\textbf {\bibinfo {volume} {316}},\ \bibinfo {pages} {1007}
	(\bibinfo {year} {2007})}\BibitemShut {NoStop}%
\bibitem [{\citenamefont {Sanvitto}\ \emph {et~al.}(2011)\citenamefont
	{Sanvitto}, \citenamefont {Pigeon}, \citenamefont {Amo}, \citenamefont
	{Ballarini}, \citenamefont {Giorgi}, \citenamefont {Carusotto}, \citenamefont
	{Hivet}, \citenamefont {Pisanello}, \citenamefont {Sala}, \citenamefont
	{Guimaraes}, \citenamefont {Houdré}, \citenamefont {Giacobino},
	\citenamefont {Ciuti}, \citenamefont {Bramati},\ and\ \citenamefont
	{Gigli}}]{sanvitto_all-optical_2011}%
\BibitemOpen
\bibfield  {author} {\bibinfo {author} {\bibfnamefont {D.}~\bibnamefont
		{Sanvitto}}, \bibinfo {author} {\bibfnamefont {S.}~\bibnamefont {Pigeon}},
	\bibinfo {author} {\bibfnamefont {A.}~\bibnamefont {Amo}}, \bibinfo {author}
	{\bibfnamefont {D.}~\bibnamefont {Ballarini}}, \bibinfo {author}
	{\bibfnamefont {M.~D.}\ \bibnamefont {Giorgi}}, \bibinfo {author}
	{\bibfnamefont {I.}~\bibnamefont {Carusotto}}, \bibinfo {author}
	{\bibfnamefont {R.}~\bibnamefont {Hivet}}, \bibinfo {author} {\bibfnamefont
		{F.}~\bibnamefont {Pisanello}}, \bibinfo {author} {\bibfnamefont {V.~G.}\
		\bibnamefont {Sala}}, \bibinfo {author} {\bibfnamefont {P.~S.~S.}\
		\bibnamefont {Guimaraes}}, \bibinfo {author} {\bibfnamefont {R.}~\bibnamefont
		{Houdré}}, \bibinfo {author} {\bibfnamefont {E.}~\bibnamefont {Giacobino}},
	\bibinfo {author} {\bibfnamefont {C.}~\bibnamefont {Ciuti}}, \bibinfo
	{author} {\bibfnamefont {A.}~\bibnamefont {Bramati}}, \ and\ \bibinfo
	{author} {\bibfnamefont {G.}~\bibnamefont {Gigli}},\ }\href {\doibase
	10.1038/nphoton.2011.211} {\bibfield  {journal} {\bibinfo  {journal} {Nature
			Photonics}\ }\textbf {\bibinfo {volume} {5}},\ \bibinfo {pages} {610}
	(\bibinfo {year} {2011})}\BibitemShut {NoStop}%
\bibitem [{\citenamefont {Askitopoulos}\ \emph {et~al.}(2013)\citenamefont
	{Askitopoulos}, \citenamefont {Ohadi}, \citenamefont {Kavokin}, \citenamefont
	{Hatzopoulos}, \citenamefont {Savvidis},\ and\ \citenamefont
	{Lagoudakis}}]{Ring}%
\BibitemOpen
\bibfield  {author} {\bibinfo {author} {\bibfnamefont {A.}~\bibnamefont
		{Askitopoulos}}, \bibinfo {author} {\bibfnamefont {H.}~\bibnamefont {Ohadi}},
	\bibinfo {author} {\bibfnamefont {A.~V.}\ \bibnamefont {Kavokin}}, \bibinfo
	{author} {\bibfnamefont {Z.}~\bibnamefont {Hatzopoulos}}, \bibinfo {author}
	{\bibfnamefont {P.~G.}\ \bibnamefont {Savvidis}}, \ and\ \bibinfo {author}
	{\bibfnamefont {P.~G.}\ \bibnamefont {Lagoudakis}},\ }\href {\doibase
	10.1103/PhysRevB.88.041308} {\bibfield  {journal} {\bibinfo  {journal}
		{Physical Review B}\ }\textbf {\bibinfo {volume} {88}},\ \bibinfo {pages}
	{041308} (\bibinfo {year} {2013})}\BibitemShut {NoStop}%
\bibitem [{\citenamefont {Dall}\ \emph {et~al.}(2014)\citenamefont {Dall},
	\citenamefont {Fraser}, \citenamefont {Desyatnikov}, \citenamefont {Li},
	\citenamefont {Brodbeck}, \citenamefont {Kamp}, \citenamefont {Schneider},
	\citenamefont {Höfling},\ and\ \citenamefont {Ostrovskaya}}]{chiral2014}%
\BibitemOpen
\bibfield  {author} {\bibinfo {author} {\bibfnamefont {R.}~\bibnamefont
		{Dall}}, \bibinfo {author} {\bibfnamefont {M.~D.}\ \bibnamefont {Fraser}},
	\bibinfo {author} {\bibfnamefont {A.~S.}\ \bibnamefont {Desyatnikov}},
	\bibinfo {author} {\bibfnamefont {G.}~\bibnamefont {Li}}, \bibinfo {author}
	{\bibfnamefont {S.}~\bibnamefont {Brodbeck}}, \bibinfo {author}
	{\bibfnamefont {M.}~\bibnamefont {Kamp}}, \bibinfo {author} {\bibfnamefont
		{C.}~\bibnamefont {Schneider}}, \bibinfo {author} {\bibfnamefont
		{S.}~\bibnamefont {Höfling}}, \ and\ \bibinfo {author} {\bibfnamefont
		{E.~A.}\ \bibnamefont {Ostrovskaya}},\ }\href {\doibase
	10.1103/PhysRevLett.113.200404} {\bibfield  {journal} {\bibinfo  {journal}
		{Physical Review Letters}\ }\textbf {\bibinfo {volume} {113}},\ \bibinfo
	{pages} {200404} (\bibinfo {year} {2014})}\BibitemShut {NoStop}%
\bibitem [{\citenamefont {Gao}\ \emph {et~al.}(2012)\citenamefont {Gao},
	\citenamefont {Eldridge}, \citenamefont {Liew}, \citenamefont {Tsintzos},
	\citenamefont {Stavrinidis}, \citenamefont {Deligeorgis}, \citenamefont
	{Hatzopoulos},\ and\ \citenamefont {Savvidis}}]{gao_polariton_2012}%
\BibitemOpen
\bibfield  {author} {\bibinfo {author} {\bibfnamefont {T.}~\bibnamefont
		{Gao}}, \bibinfo {author} {\bibfnamefont {P.~S.}\ \bibnamefont {Eldridge}},
	\bibinfo {author} {\bibfnamefont {T.~C.~H.}\ \bibnamefont {Liew}}, \bibinfo
	{author} {\bibfnamefont {S.~I.}\ \bibnamefont {Tsintzos}}, \bibinfo {author}
	{\bibfnamefont {G.}~\bibnamefont {Stavrinidis}}, \bibinfo {author}
	{\bibfnamefont {G.}~\bibnamefont {Deligeorgis}}, \bibinfo {author}
	{\bibfnamefont {Z.}~\bibnamefont {Hatzopoulos}}, \ and\ \bibinfo {author}
	{\bibfnamefont {P.~G.}\ \bibnamefont {Savvidis}},\ }\href {\doibase
	10.1103/PhysRevB.85.235102} {\bibfield  {journal} {\bibinfo  {journal}
		{Physical Review B}\ }\textbf {\bibinfo {volume} {85}},\ \bibinfo {pages}
	{235102} (\bibinfo {year} {2012})}\BibitemShut {NoStop}%
\bibitem [{\citenamefont {Manni}\ \emph {et~al.}(2011)\citenamefont {Manni},
	\citenamefont {Lagoudakis}, \citenamefont {Liew}, \citenamefont {André},\
	and\ \citenamefont {Deveaud-Plédran}}]{manni_spontaneous_2011}%
\BibitemOpen
\bibfield  {author} {\bibinfo {author} {\bibfnamefont {F.}~\bibnamefont
		{Manni}}, \bibinfo {author} {\bibfnamefont {K.~G.}\ \bibnamefont
		{Lagoudakis}}, \bibinfo {author} {\bibfnamefont {T.~C.~H.}\ \bibnamefont
		{Liew}}, \bibinfo {author} {\bibfnamefont {R.}~\bibnamefont {André}}, \ and\
	\bibinfo {author} {\bibfnamefont {B.}~\bibnamefont {Deveaud-Plédran}},\
}\href {\doibase 10.1103/PhysRevLett.107.106401} {\bibfield  {journal}
{\bibinfo  {journal} {Physical Review Letters}\ }\textbf {\bibinfo {volume}
	{107}},\ \bibinfo {pages} {106401} (\bibinfo {year} {2011})}\BibitemShut
{NoStop}%
\bibitem [{\citenamefont {Cristofolini}\ \emph {et~al.}(2013)\citenamefont
	{Cristofolini}, \citenamefont {Dreismann}, \citenamefont {Christmann},
	\citenamefont {Franchetti}, \citenamefont {Berloff}, \citenamefont {Tsotsis},
	\citenamefont {Hatzopoulos}, \citenamefont {Savvidis},\ and\ \citenamefont
	{Baumberg}}]{opticalSPT2013}%
\BibitemOpen
\bibfield  {author} {\bibinfo {author} {\bibfnamefont {P.}~\bibnamefont
		{Cristofolini}}, \bibinfo {author} {\bibfnamefont {A.}~\bibnamefont
		{Dreismann}}, \bibinfo {author} {\bibfnamefont {G.}~\bibnamefont
		{Christmann}}, \bibinfo {author} {\bibfnamefont {G.}~\bibnamefont
		{Franchetti}}, \bibinfo {author} {\bibfnamefont {N.~G.}\ \bibnamefont
		{Berloff}}, \bibinfo {author} {\bibfnamefont {P.}~\bibnamefont {Tsotsis}},
	\bibinfo {author} {\bibfnamefont {Z.}~\bibnamefont {Hatzopoulos}}, \bibinfo
	{author} {\bibfnamefont {P.~G.}\ \bibnamefont {Savvidis}}, \ and\ \bibinfo
	{author} {\bibfnamefont {J.~J.}\ \bibnamefont {Baumberg}},\ }\href {\doibase
	10.1103/PhysRevLett.110.186403} {\bibfield  {journal} {\bibinfo  {journal}
		{Physical Review Letters}\ }\textbf {\bibinfo {volume} {110}},\ \bibinfo
	{pages} {186403} (\bibinfo {year} {2013})}\BibitemShut {NoStop}%
\bibitem [{\citenamefont {Dreismann}\ \emph {et~al.}(2014)\citenamefont
	{Dreismann}, \citenamefont {Cristofolini}, \citenamefont {Balili},
	\citenamefont {Christmann}, \citenamefont {Pinsker}, \citenamefont {Berloff},
	\citenamefont {Hatzopoulos}, \citenamefont {Savvidis},\ and\ \citenamefont
	{Baumberg}}]{rotating_2014}%
\BibitemOpen
\bibfield  {author} {\bibinfo {author} {\bibfnamefont {A.}~\bibnamefont
		{Dreismann}}, \bibinfo {author} {\bibfnamefont {P.}~\bibnamefont
		{Cristofolini}}, \bibinfo {author} {\bibfnamefont {R.}~\bibnamefont
		{Balili}}, \bibinfo {author} {\bibfnamefont {G.}~\bibnamefont {Christmann}},
	\bibinfo {author} {\bibfnamefont {F.}~\bibnamefont {Pinsker}}, \bibinfo
	{author} {\bibfnamefont {N.~G.}\ \bibnamefont {Berloff}}, \bibinfo {author}
	{\bibfnamefont {Z.}~\bibnamefont {Hatzopoulos}}, \bibinfo {author}
	{\bibfnamefont {P.~G.}\ \bibnamefont {Savvidis}}, \ and\ \bibinfo {author}
	{\bibfnamefont {J.~J.}\ \bibnamefont {Baumberg}},\ }\href {\doibase
	10.1073/pnas.1401988111} {\bibfield  {journal} {\bibinfo  {journal}
		{Proceedings of the National Academy of Sciences}\ ,\ \bibinfo {pages}
		{201401988}} (\bibinfo {year} {2014})}\BibitemShut {NoStop}%
\bibitem [{\citenamefont {Kalevich}\ \emph {et~al.}(2014)\citenamefont
	{Kalevich}, \citenamefont {Afanasiev}, \citenamefont {Lukoshkin},
	\citenamefont {Kavokin}, \citenamefont {Tsintzos}, \citenamefont {Savvidis},\
	and\ \citenamefont {Kavokin}}]{kalevich}%
\BibitemOpen
\bibfield  {author} {\bibinfo {author} {\bibfnamefont {V.~K.}\ \bibnamefont
		{Kalevich}}, \bibinfo {author} {\bibfnamefont {M.~M.}\ \bibnamefont
		{Afanasiev}}, \bibinfo {author} {\bibfnamefont {V.~A.}\ \bibnamefont
		{Lukoshkin}}, \bibinfo {author} {\bibfnamefont {K.~V.}\ \bibnamefont
		{Kavokin}}, \bibinfo {author} {\bibfnamefont {S.~I.}\ \bibnamefont
		{Tsintzos}}, \bibinfo {author} {\bibfnamefont {P.~G.}\ \bibnamefont
		{Savvidis}}, \ and\ \bibinfo {author} {\bibfnamefont {A.~V.}\ \bibnamefont
		{Kavokin}},\ }\href {\doibase 10.1063/1.4867519} {\bibfield  {journal}
	{\bibinfo  {journal} {Journal of Applied Physics}\ }\textbf {\bibinfo
		{volume} {115}},\ \bibinfo {pages} {094304} (\bibinfo {year}
	{2014})}\BibitemShut {NoStop}%
\bibitem [{\citenamefont {Ohadi}\ \emph {et~al.}(2014)\citenamefont {Ohadi},
	\citenamefont {Gregory}, \citenamefont {Freegarde}, \citenamefont {Rubo},
	\citenamefont {Kavokin},\ and\ \citenamefont
	{Lagoudakis}}]{ohadi_dissipative_2014}%
\BibitemOpen
\bibfield  {author} {\bibinfo {author} {\bibfnamefont {H.}~\bibnamefont
		{Ohadi}}, \bibinfo {author} {\bibfnamefont {R.~L.}\ \bibnamefont {Gregory}},
	\bibinfo {author} {\bibfnamefont {T.}~\bibnamefont {Freegarde}}, \bibinfo
	{author} {\bibfnamefont {Y.~G.}\ \bibnamefont {Rubo}}, \bibinfo {author}
	{\bibfnamefont {A.~V.}\ \bibnamefont {Kavokin}}, \ and\ \bibinfo {author}
	{\bibfnamefont {P.~G.}\ \bibnamefont {Lagoudakis}},\ }\href
{http://arxiv.org/abs/1406.6377} {\bibfield  {journal} {\bibinfo  {journal}
		{{arXiv}:1406.6377 [cond-mat]}\ } (\bibinfo {year} {2014})},\ \bibinfo {note}
{{arXiv}: 1406.6377}\BibitemShut {NoStop}%
\bibitem [{\citenamefont {Schmutzler}\ \emph {et~al.}(2015)\citenamefont
	{Schmutzler}, \citenamefont {Lewandowski}, \citenamefont {Aßmann},
	\citenamefont {Niemietz}, \citenamefont {Schumacher}, \citenamefont {Kamp},
	\citenamefont {Schneider}, \citenamefont {Höfling},\ and\ \citenamefont
	{Bayer}}]{schmutzler_all-optical_2015}%
\BibitemOpen
\bibfield  {author} {\bibinfo {author} {\bibfnamefont {J.}~\bibnamefont
		{Schmutzler}}, \bibinfo {author} {\bibfnamefont {P.}~\bibnamefont
		{Lewandowski}}, \bibinfo {author} {\bibfnamefont {M.}~\bibnamefont
		{Aßmann}}, \bibinfo {author} {\bibfnamefont {D.}~\bibnamefont {Niemietz}},
	\bibinfo {author} {\bibfnamefont {S.}~\bibnamefont {Schumacher}}, \bibinfo
	{author} {\bibfnamefont {M.}~\bibnamefont {Kamp}}, \bibinfo {author}
	{\bibfnamefont {C.}~\bibnamefont {Schneider}}, \bibinfo {author}
	{\bibfnamefont {S.}~\bibnamefont {Höfling}}, \ and\ \bibinfo {author}
	{\bibfnamefont {M.}~\bibnamefont {Bayer}},\ }\href {\doibase
	10.1103/PhysRevB.91.195308} {\bibfield  {journal} {\bibinfo  {journal}
		{Physical Review B}\ }\textbf {\bibinfo {volume} {91}},\ \bibinfo {pages}
	{195308} (\bibinfo {year} {2015})}\BibitemShut {NoStop}%
\bibitem [{\citenamefont {Byrnes}\ \emph {et~al.}(2014)\citenamefont {Byrnes},
	\citenamefont {Kim},\ and\ \citenamefont {Yamamoto}}]{YamamotoReview_2014}%
\BibitemOpen
\bibfield  {author} {\bibinfo {author} {\bibfnamefont {T.}~\bibnamefont
		{Byrnes}}, \bibinfo {author} {\bibfnamefont {N.~Y.}\ \bibnamefont {Kim}}, \
	and\ \bibinfo {author} {\bibfnamefont {Y.}~\bibnamefont {Yamamoto}},\ }\href
{\doibase 10.1038/nphys3143} {\bibfield  {journal} {\bibinfo  {journal}
		{Nature Physics}\ }\textbf {\bibinfo {volume} {10}},\ \bibinfo {pages} {803}
	(\bibinfo {year} {2014})}\BibitemShut {NoStop}%
\bibitem [{\citenamefont {Demirchyan}\ \emph {et~al.}(2014)\citenamefont
	{Demirchyan}, \citenamefont {Chestnov}, \citenamefont {Alodjants},
	\citenamefont {Glazov},\ and\ \citenamefont {Kavokin}}]{qubits_2014}%
\BibitemOpen
\bibfield  {author} {\bibinfo {author} {\bibfnamefont {S.}~\bibnamefont
		{Demirchyan}}, \bibinfo {author} {\bibfnamefont {I.}~\bibnamefont
		{Chestnov}}, \bibinfo {author} {\bibfnamefont {A.}~\bibnamefont {Alodjants}},
	\bibinfo {author} {\bibfnamefont {M.}~\bibnamefont {Glazov}}, \ and\ \bibinfo
	{author} {\bibfnamefont {A.}~\bibnamefont {Kavokin}},\ }\href {\doibase
	10.1103/PhysRevLett.112.196403} {\bibfield  {journal} {\bibinfo  {journal}
		{Physical Review Letters}\ }\textbf {\bibinfo {volume} {112}},\ \bibinfo
	{pages} {196403} (\bibinfo {year} {2014})}\BibitemShut {NoStop}%
\bibitem [{\citenamefont {Sanvitto}\ \emph {et~al.}(2009)\citenamefont
	{Sanvitto}, \citenamefont {Amo}, \citenamefont {Viña}, \citenamefont
	{André}, \citenamefont {Solnyshkov},\ and\ \citenamefont
	{Malpuech}}]{sanvitto2009}%
\BibitemOpen
\bibfield  {author} {\bibinfo {author} {\bibfnamefont {D.}~\bibnamefont
		{Sanvitto}}, \bibinfo {author} {\bibfnamefont {A.}~\bibnamefont {Amo}},
	\bibinfo {author} {\bibfnamefont {L.}~\bibnamefont {Viña}}, \bibinfo
	{author} {\bibfnamefont {R.}~\bibnamefont {André}}, \bibinfo {author}
	{\bibfnamefont {D.}~\bibnamefont {Solnyshkov}}, \ and\ \bibinfo {author}
	{\bibfnamefont {G.}~\bibnamefont {Malpuech}},\ }\href {\doibase
	10.1103/PhysRevB.80.045301} {\bibfield  {journal} {\bibinfo  {journal}
		{Physical Review B}\ }\textbf {\bibinfo {volume} {80}},\ \bibinfo {pages}
	{045301} (\bibinfo {year} {2009})}\BibitemShut {NoStop}%
\bibitem [{\citenamefont {Maragkou}\ \emph {et~al.}(2010)\citenamefont
	{Maragkou}, \citenamefont {Grundy}, \citenamefont {Wertz}, \citenamefont
	{Lemaître}, \citenamefont {Sagnes}, \citenamefont {Senellart}, \citenamefont
	{Bloch},\ and\ \citenamefont {Lagoudakis}}]{maragkou}%
\BibitemOpen
\bibfield  {author} {\bibinfo {author} {\bibfnamefont {M.}~\bibnamefont
		{Maragkou}}, \bibinfo {author} {\bibfnamefont {A.~J.~D.}\ \bibnamefont
		{Grundy}}, \bibinfo {author} {\bibfnamefont {E.}~\bibnamefont {Wertz}},
	\bibinfo {author} {\bibfnamefont {A.}~\bibnamefont {Lemaître}}, \bibinfo
	{author} {\bibfnamefont {I.}~\bibnamefont {Sagnes}}, \bibinfo {author}
	{\bibfnamefont {P.}~\bibnamefont {Senellart}}, \bibinfo {author}
	{\bibfnamefont {J.}~\bibnamefont {Bloch}}, \ and\ \bibinfo {author}
	{\bibfnamefont {P.~G.}\ \bibnamefont {Lagoudakis}},\ }\href {\doibase
	10.1103/PhysRevB.81.081307} {\bibfield  {journal} {\bibinfo  {journal}
		{Physical Review B}\ }\textbf {\bibinfo {volume} {81}},\ \bibinfo {pages}
	{081307} (\bibinfo {year} {2010})}\BibitemShut {NoStop}%
\bibitem [{\citenamefont {Nardin}\ \emph {et~al.}(2010)\citenamefont {Nardin},
	\citenamefont {Léger}, \citenamefont {Pietka}, \citenamefont
	{Morier-Genoud},\ and\ \citenamefont {Deveaud-Plédran}}]{nardin_2010}%
\BibitemOpen
\bibfield  {author} {\bibinfo {author} {\bibfnamefont {G.}~\bibnamefont
		{Nardin}}, \bibinfo {author} {\bibfnamefont {Y.}~\bibnamefont {Léger}},
	\bibinfo {author} {\bibfnamefont {B.}~\bibnamefont {Pietka}}, \bibinfo
	{author} {\bibfnamefont {F.}~\bibnamefont {Morier-Genoud}}, \ and\ \bibinfo
	{author} {\bibfnamefont {B.}~\bibnamefont {Deveaud-Plédran}},\ }\href
{\doibase 10.1103/PhysRevB.82.045304} {\bibfield  {journal} {\bibinfo
		{journal} {Physical Review B}\ }\textbf {\bibinfo {volume} {82}},\ \bibinfo
	{pages} {045304} (\bibinfo {year} {2010})}\BibitemShut {NoStop}%
\bibitem [{\citenamefont {Eastham}(2008)}]{eastham2008}%
\BibitemOpen
\bibfield  {author} {\bibinfo {author} {\bibfnamefont {P.~R.}\ \bibnamefont
		{Eastham}},\ }\href {\doibase 10.1103/PhysRevB.78.035319} {\bibfield
	{journal} {\bibinfo  {journal} {Physical Review B}\ }\textbf {\bibinfo
		{volume} {78}},\ \bibinfo {pages} {035319} (\bibinfo {year}
	{2008})}\BibitemShut {NoStop}%
\bibitem [{\citenamefont {Portolan}\ \emph {et~al.}(2008)\citenamefont
	{Portolan}, \citenamefont {Hauke},\ and\ \citenamefont
	{Savona}}]{portolan2008}%
\BibitemOpen
\bibfield  {author} {\bibinfo {author} {\bibfnamefont {S.}~\bibnamefont
		{Portolan}}, \bibinfo {author} {\bibfnamefont {P.}~\bibnamefont {Hauke}}, \
	and\ \bibinfo {author} {\bibfnamefont {V.}~\bibnamefont {Savona}},\ }\href
{\doibase 10.1002/pssb.200777626} {\bibfield  {journal} {\bibinfo  {journal}
		{physica status solidi (b)}\ }\textbf {\bibinfo {volume} {245}},\ \bibinfo
	{pages} {1089} (\bibinfo {year} {2008})}\BibitemShut {NoStop}%
\bibitem [{\citenamefont {Tsotsis}\ \emph {et~al.}(2012)\citenamefont
	{Tsotsis}, \citenamefont {Eldridge}, \citenamefont {Gao}, \citenamefont
	{Tsintzos}, \citenamefont {Hatzopoulos},\ and\ \citenamefont
	{Savvidis}}]{tsotsis_lasing_2012}%
\BibitemOpen
\bibfield  {author} {\bibinfo {author} {\bibfnamefont {P.}~\bibnamefont
		{Tsotsis}}, \bibinfo {author} {\bibfnamefont {P.~S.}\ \bibnamefont
		{Eldridge}}, \bibinfo {author} {\bibfnamefont {T.}~\bibnamefont {Gao}},
	\bibinfo {author} {\bibfnamefont {S.~I.}\ \bibnamefont {Tsintzos}}, \bibinfo
	{author} {\bibfnamefont {Z.}~\bibnamefont {Hatzopoulos}}, \ and\ \bibinfo
	{author} {\bibfnamefont {P.~G.}\ \bibnamefont {Savvidis}},\ }\href {\doibase
	10.1088/1367-2630/14/2/023060} {\bibfield  {journal} {\bibinfo  {journal}
		{New Journal of Physics}\ }\textbf {\bibinfo {volume} {14}},\ \bibinfo
	{pages} {023060} (\bibinfo {year} {2012})}\BibitemShut {NoStop}%
\bibitem [{sup()}]{supp}%
\BibitemOpen
\href@noop {} {}\bibinfo {note} {See supllementary information accompanying
	this paper}\BibitemShut {NoStop}%
\bibitem [{\citenamefont {Wouters}\ and\ \citenamefont
	{Carusotto}(2007)}]{Wouters2007}%
\BibitemOpen
\bibfield  {author} {\bibinfo {author} {\bibfnamefont {M.}~\bibnamefont
		{Wouters}}\ and\ \bibinfo {author} {\bibfnamefont {I.}~\bibnamefont
		{Carusotto}},\ }\href {\doibase 10.1103/PhysRevLett.99.140402} {\bibfield
	{journal} {\bibinfo  {journal} {Physical Review Letters}\ }\textbf {\bibinfo
		{volume} {99}},\ \bibinfo {pages} {140402} (\bibinfo {year}
	{2007})}\BibitemShut {NoStop}%
\bibitem [{\citenamefont {Keeling}\ and\ \citenamefont
	{Berloff}(2008)}]{Keeling2008}%
\BibitemOpen
\bibfield  {author} {\bibinfo {author} {\bibfnamefont {J.}~\bibnamefont
		{Keeling}}\ and\ \bibinfo {author} {\bibfnamefont {N.~G.}\ \bibnamefont
		{Berloff}},\ }\href {\doibase 10.1103/PhysRevLett.100.250401} {\bibfield
	{journal} {\bibinfo  {journal} {Physical Review Letters}\ }\textbf {\bibinfo
		{volume} {100}},\ \bibinfo {pages} {250401} (\bibinfo {year}
	{2008})}\BibitemShut {NoStop}%
\bibitem [{\citenamefont {Wouters}(2012)}]{Wouters2012}%
\BibitemOpen
\bibfield  {author} {\bibinfo {author} {\bibfnamefont {M.}~\bibnamefont
		{Wouters}},\ }\href {\doibase 10.1088/1367-2630/14/7/075020} {\bibfield
	{journal} {\bibinfo  {journal} {New Journal of Physics}\ }\textbf {\bibinfo
		{volume} {14}},\ \bibinfo {pages} {075020} (\bibinfo {year}
	{2012})}\BibitemShut {NoStop}%
\bibitem [{\citenamefont {Wouters}\ \emph {et~al.}(2010)\citenamefont
	{Wouters}, \citenamefont {Liew},\ and\ \citenamefont {Savona}}]{Wouters2010}%
\BibitemOpen
\bibfield  {author} {\bibinfo {author} {\bibfnamefont {M.}~\bibnamefont
		{Wouters}}, \bibinfo {author} {\bibfnamefont {T.~C.~H.}\ \bibnamefont
		{Liew}}, \ and\ \bibinfo {author} {\bibfnamefont {V.}~\bibnamefont
		{Savona}},\ }\href {\doibase 10.1103/PhysRevB.82.245315} {\bibfield
	{journal} {\bibinfo  {journal} {Physical Review B}\ }\textbf {\bibinfo
		{volume} {82}},\ \bibinfo {pages} {245315} (\bibinfo {year}
	{2010})}\BibitemShut {NoStop}%
\bibitem [{\citenamefont {Wertz}\ \emph {et~al.}(2012)\citenamefont {Wertz},
	\citenamefont {Amo}, \citenamefont {Solnyshkov}, \citenamefont {Ferrier},
	\citenamefont {Liew}, \citenamefont {Sanvitto}, \citenamefont {Senellart},
	\citenamefont {Sagnes}, \citenamefont {Lemaître}, \citenamefont {Kavokin},
	\citenamefont {Malpuech},\ and\ \citenamefont {Bloch}}]{Wertz2012}%
\BibitemOpen
\bibfield  {author} {\bibinfo {author} {\bibfnamefont {E.}~\bibnamefont
		{Wertz}}, \bibinfo {author} {\bibfnamefont {A.}~\bibnamefont {Amo}}, \bibinfo
	{author} {\bibfnamefont {D.~D.}\ \bibnamefont {Solnyshkov}}, \bibinfo
	{author} {\bibfnamefont {L.}~\bibnamefont {Ferrier}}, \bibinfo {author}
	{\bibfnamefont {T.~C.~H.}\ \bibnamefont {Liew}}, \bibinfo {author}
	{\bibfnamefont {D.}~\bibnamefont {Sanvitto}}, \bibinfo {author}
	{\bibfnamefont {P.}~\bibnamefont {Senellart}}, \bibinfo {author}
	{\bibfnamefont {I.}~\bibnamefont {Sagnes}}, \bibinfo {author} {\bibfnamefont
		{A.}~\bibnamefont {Lemaître}}, \bibinfo {author} {\bibfnamefont {A.~V.}\
		\bibnamefont {Kavokin}}, \bibinfo {author} {\bibfnamefont {G.}~\bibnamefont
		{Malpuech}}, \ and\ \bibinfo {author} {\bibfnamefont {J.}~\bibnamefont
		{Bloch}},\ }\href {\doibase 10.1103/PhysRevLett.109.216404} {\bibfield
	{journal} {\bibinfo  {journal} {Physical Review Letters}\ }\textbf {\bibinfo
		{volume} {109}},\ \bibinfo {pages} {216404} (\bibinfo {year}
	{2012})}\BibitemShut {NoStop}%
\bibitem [{\citenamefont {Antón}\ \emph {et~al.}(2012)\citenamefont {Antón},
	\citenamefont {Liew}, \citenamefont {Tosi}, \citenamefont {Martín},
	\citenamefont {Gao}, \citenamefont {Hatzopoulos}, \citenamefont {Eldridge},
	\citenamefont {Savvidis},\ and\ \citenamefont {Viña}}]{Anton2012}%
\BibitemOpen
\bibfield  {author} {\bibinfo {author} {\bibfnamefont {C.}~\bibnamefont
		{Antón}}, \bibinfo {author} {\bibfnamefont {T.~C.~H.}\ \bibnamefont {Liew}},
	\bibinfo {author} {\bibfnamefont {G.}~\bibnamefont {Tosi}}, \bibinfo {author}
	{\bibfnamefont {M.~D.}\ \bibnamefont {Martín}}, \bibinfo {author}
	{\bibfnamefont {T.}~\bibnamefont {Gao}}, \bibinfo {author} {\bibfnamefont
		{Z.}~\bibnamefont {Hatzopoulos}}, \bibinfo {author} {\bibfnamefont {P.~S.}\
		\bibnamefont {Eldridge}}, \bibinfo {author} {\bibfnamefont {P.~G.}\
		\bibnamefont {Savvidis}}, \ and\ \bibinfo {author} {\bibfnamefont
		{L.}~\bibnamefont {Viña}},\ }\href {\doibase 10.1063/1.4773376} {\bibfield
	{journal} {\bibinfo  {journal} {Applied Physics Letters}\ }\textbf {\bibinfo
		{volume} {101}},\ \bibinfo {pages} {261116} (\bibinfo {year}
	{2012})}\BibitemShut {NoStop}%
\bibitem [{\citenamefont {Solnyshkov}\ \emph {et~al.}(2014)\citenamefont
	{Solnyshkov}, \citenamefont {Terças}, \citenamefont {Dini},\ and\
	\citenamefont {Malpuech}}]{Solnyshkov2014}%
\BibitemOpen
\bibfield  {author} {\bibinfo {author} {\bibfnamefont {D.~D.}\ \bibnamefont
		{Solnyshkov}}, \bibinfo {author} {\bibfnamefont {H.}~\bibnamefont {Terças}},
	\bibinfo {author} {\bibfnamefont {K.}~\bibnamefont {Dini}}, \ and\ \bibinfo
	{author} {\bibfnamefont {G.}~\bibnamefont {Malpuech}},\ }\href {\doibase
	10.1103/PhysRevA.89.033626} {\bibfield  {journal} {\bibinfo  {journal}
		{Physical Review A}\ }\textbf {\bibinfo {volume} {89}},\ \bibinfo {pages}
	{033626} (\bibinfo {year} {2014})}\BibitemShut {NoStop}%
\bibitem [{\citenamefont {Sieberer}\ \emph {et~al.}(2014)\citenamefont
	{Sieberer}, \citenamefont {Huber}, \citenamefont {Altman},\ and\
	\citenamefont {Diehl}}]{Sieberer2014}%
\BibitemOpen
\bibfield  {author} {\bibinfo {author} {\bibfnamefont {L.~M.}\ \bibnamefont
		{Sieberer}}, \bibinfo {author} {\bibfnamefont {S.~D.}\ \bibnamefont {Huber}},
	\bibinfo {author} {\bibfnamefont {E.}~\bibnamefont {Altman}}, \ and\ \bibinfo
	{author} {\bibfnamefont {S.}~\bibnamefont {Diehl}},\ }\href {\doibase
	10.1103/PhysRevB.89.134310} {\bibfield  {journal} {\bibinfo  {journal}
		{Physical Review B}\ }\textbf {\bibinfo {volume} {89}},\ \bibinfo {pages}
	{134310} (\bibinfo {year} {2014})}\BibitemShut {NoStop}%
\bibitem [{\citenamefont {Haug}\ \emph {et~al.}(2014)\citenamefont {Haug},
	\citenamefont {Doan},\ and\ \citenamefont {Tran~Thoai}}]{Haug2014}%
\BibitemOpen
\bibfield  {author} {\bibinfo {author} {\bibfnamefont {H.}~\bibnamefont
		{Haug}}, \bibinfo {author} {\bibfnamefont {T.~D.}\ \bibnamefont {Doan}}, \
	and\ \bibinfo {author} {\bibfnamefont {D.~B.}\ \bibnamefont {Tran~Thoai}},\
}\href {\doibase 10.1103/PhysRevB.89.155302} {\bibfield  {journal} {\bibinfo
	{journal} {Physical Review B}\ }\textbf {\bibinfo {volume} {89}},\ \bibinfo
{pages} {155302} (\bibinfo {year} {2014})}\BibitemShut {NoStop}%
\bibitem [{Tim()}]{TimV}%
\BibitemOpen
\href@noop {} {}\bibinfo {note} {We used the following parameters to describe
	our system: $m_P=7\times10^{-5}m_e$,
	$U_{pol-pol}=2.4\times10^{-3}$~meV$\mu$m$^2$, $U_{pol-ex}=2U_{pol-pol}$,
	$\Gamma_\mathrm{NL}=0.3 U_{pol-pol}$~\cite{Keeling2008},
	$r=6\times10^{-4}$meV$\mu$m$^2$,
	$\lambda=1.2\times10^{-3}\mu$m$^2$ps$^{-1}$meV$^{-1}$,
	$\hbar/\Gamma=5$ps.}\BibitemShut {Stop}%
\bibitem [{\citenamefont {Combescot}\ and\ \citenamefont
	{Snoke}(2008)}]{combescot_stability_2008}%
\BibitemOpen
\bibfield  {author} {\bibinfo {author} {\bibfnamefont {M.}~\bibnamefont
		{Combescot}}\ and\ \bibinfo {author} {\bibfnamefont {D.~W.}\ \bibnamefont
		{Snoke}},\ }\href {\doibase 10.1103/PhysRevB.78.144303} {\bibfield  {journal}
	{\bibinfo  {journal} {Physical Review B}\ }\textbf {\bibinfo {volume} {78}},\
	\bibinfo {pages} {144303} (\bibinfo {year} {2008})}\BibitemShut {NoStop}%
\bibitem [{\citenamefont {Nelsen}\ \emph {et~al.}(2013)\citenamefont {Nelsen},
	\citenamefont {Liu}, \citenamefont {Steger}, \citenamefont {Snoke},
	\citenamefont {Balili}, \citenamefont {West},\ and\ \citenamefont
	{Pfeiffer}}]{nelsen_dissipationless_2013}%
\BibitemOpen
\bibfield  {author} {\bibinfo {author} {\bibfnamefont {B.}~\bibnamefont
		{Nelsen}}, \bibinfo {author} {\bibfnamefont {G.}~\bibnamefont {Liu}},
	\bibinfo {author} {\bibfnamefont {M.}~\bibnamefont {Steger}}, \bibinfo
	{author} {\bibfnamefont {D.~W.}\ \bibnamefont {Snoke}}, \bibinfo {author}
	{\bibfnamefont {R.}~\bibnamefont {Balili}}, \bibinfo {author} {\bibfnamefont
		{K.}~\bibnamefont {West}}, \ and\ \bibinfo {author} {\bibfnamefont
		{L.}~\bibnamefont {Pfeiffer}},\ }\href {\doibase 10.1103/PhysRevX.3.041015}
{\bibfield  {journal} {\bibinfo  {journal} {Physical Review X}\ }\textbf
	{\bibinfo {volume} {3}},\ \bibinfo {pages} {041015} (\bibinfo {year}
	{2013})}\BibitemShut {NoStop}%
\bibitem [{\citenamefont {Liew}\ \emph {et~al.}(2013)\citenamefont {Liew},
	\citenamefont {Glazov}, \citenamefont {Kavokin}, \citenamefont {Shelykh},
	\citenamefont {Kaliteevski},\ and\ \citenamefont
	{Kavokin}}]{liew_proposal_2013}%
\BibitemOpen
\bibfield  {author} {\bibinfo {author} {\bibfnamefont {T.~C.~H.}\
		\bibnamefont {Liew}}, \bibinfo {author} {\bibfnamefont {M.~M.}\ \bibnamefont
		{Glazov}}, \bibinfo {author} {\bibfnamefont {K.~V.}\ \bibnamefont {Kavokin}},
	\bibinfo {author} {\bibfnamefont {I.~A.}\ \bibnamefont {Shelykh}}, \bibinfo
	{author} {\bibfnamefont {M.~A.}\ \bibnamefont {Kaliteevski}}, \ and\ \bibinfo
	{author} {\bibfnamefont {A.~V.}\ \bibnamefont {Kavokin}},\ }\href {\doibase
	10.1103/PhysRevLett.110.047402} {\bibfield  {journal} {\bibinfo  {journal}
		{Physical Review Letters}\ }\textbf {\bibinfo {volume} {110}},\ \bibinfo
	{pages} {047402} (\bibinfo {year} {2013})}\BibitemShut {NoStop}%
\bibitem [{\citenamefont {Nguyen}\ \emph {et~al.}(2013)\citenamefont {Nguyen},
	\citenamefont {Vishnevsky}, \citenamefont {Sturm}, \citenamefont {Tanese},
	\citenamefont {Solnyshkov}, \citenamefont {Galopin}, \citenamefont
	{Lemaître}, \citenamefont {Sagnes}, \citenamefont {Amo}, \citenamefont
	{Malpuech},\ and\ \citenamefont {Bloch}}]{nguyen_realization_2013}%
\BibitemOpen
\bibfield  {author} {\bibinfo {author} {\bibfnamefont {H.~S.}\ \bibnamefont
		{Nguyen}}, \bibinfo {author} {\bibfnamefont {D.}~\bibnamefont {Vishnevsky}},
	\bibinfo {author} {\bibfnamefont {C.}~\bibnamefont {Sturm}}, \bibinfo
	{author} {\bibfnamefont {D.}~\bibnamefont {Tanese}}, \bibinfo {author}
	{\bibfnamefont {D.}~\bibnamefont {Solnyshkov}}, \bibinfo {author}
	{\bibfnamefont {E.}~\bibnamefont {Galopin}}, \bibinfo {author} {\bibfnamefont
		{A.}~\bibnamefont {Lemaître}}, \bibinfo {author} {\bibfnamefont
		{I.}~\bibnamefont {Sagnes}}, \bibinfo {author} {\bibfnamefont
		{A.}~\bibnamefont {Amo}}, \bibinfo {author} {\bibfnamefont {G.}~\bibnamefont
		{Malpuech}}, \ and\ \bibinfo {author} {\bibfnamefont {J.}~\bibnamefont
		{Bloch}},\ }\href {\doibase 10.1103/PhysRevLett.110.236601} {\bibfield
	{journal} {\bibinfo  {journal} {Physical Review Letters}\ }\textbf {\bibinfo
		{volume} {110}},\ \bibinfo {pages} {236601} (\bibinfo {year}
	{2013})}\BibitemShut {NoStop}%
\end{thebibliography}

%
\end{document}